\documentclass{article}

\usepackage{natbib}

\usepackage[]{graphicx}
\usepackage[]{color}

\definecolor{fgcolor}{rgb}{0.345, 0.345, 0.345}

\definecolor{shadecolor}{rgb}{.97, .97, .97}
\definecolor{messagecolor}{rgb}{0, 0, 0}
\definecolor{warningcolor}{rgb}{1, 0, 1}
\definecolor{errorcolor}{rgb}{1, 0, 0}

\usepackage{alltt}
\usepackage[latin1]{inputenc}
\usepackage{amsmath}
\usepackage{amsfonts}
\usepackage{amssymb}
\usepackage{amsthm}
\usepackage{graphicx}
\usepackage{bm}
\usepackage{rotating}
\usepackage[lined,boxed]{algorithm2e}

\newtheorem{property}{Property}
\newtheorem{theorem}{Theorem}

\newcommand{\ma}[1]{\ensuremath{\mathbf{#1}}}
\newcommand{\tr}{\mathrm{tr}}
\newcommand{\thetai}{\bm{\theta}_i}

\usepackage[english]{babel}

\begin{document} 

\author{Patrick J.F. Groenen and Julie Josse}
\title{Multinomial Multiple Correspondence Analysis}
\date{\today} 

\maketitle

\begin{abstract}
Relations between categorical variables can be analyzed conveniently by multiple correspondence analysis (MCA). 
The graphical representation of MCA results in so-called biplots makes it easy to interpret the most important associations. However, a major drawback of MCA is that it does not have an underlying probability model for an individual selecting a category on a variable. In this paper, we propose such probability model called multinomial multiple correspondence analysis (MMCA) that combines the underlying low-rank representation of MCA with maximum likelihood. An efficient majorization algorithm that uses an elegant bound for the second derivative is derived to estimate the parameters. The proposed model can easily lead to overfitting causing some of the parameters to wander of to infinity. We add the nuclear norm penalty to counter this issue and discuss ways of selecting regularization parameters. The proposed approach is well suited to study and vizualise the dependences for high dimensional data.
\end{abstract}

\section{Introduction}

Data sets with categorical variables are common in many fields such as social sciences, where surveys with categorical questions are conducted. Although some models are available to describe the dependence between categorical variables, they suffer from estimation issues as the number of parameters quickly grows with large number of categories and variables.  

To give a concrete example let us consider a data set from the French national institute for prevention and health education (INPES \footnote{http://www.inpes.sante.fr/default.asp}) on alcohol usage.
Each year, more than 50,000 individuals describe their consumption (kind of beverage,  frequency of  drinking,  frequency of binge drinking, etc) as well as their socio-economic and demographic characteristics.  Describing the relationships between these categorical variables is important to monitor alcohol usage in subgroups, to improve the understanding of alcohol usage, to monitor their evolution, and to suggest suitable policies.  Therefore, scientists need methods to explore such data.  

High dimensional data like this one do not fall within the scope of classical models such as the log-linear models \cite{Christensen90}. In addition, such models often lack some follow-up graphical representations which may help the user to a great extent to shed more light into the obtained results. On the other hand, principal component methods based such as multiple correspondence analysis (MCA) \cite{Green06}
are powerful techniques to explore and visualize large categorical data using biplot representation. MCA has the great advantage of being easily solved by singular value decomposition (SVD). However, MCA is often motivated by geometrical considerations without any reference to probability models.  

Our {\em multinomial multiple correspondence analysis} model aims at bringing the best from both worlds: appropriately modeling the probability of selecting a category out of several options combined with the capability of handling high dimensional data while providing graphical output to explore the relations and gain insight for interpretation. We use a parsimonious low rank representation of the data and  derive an efficient majorization algorithm to estimate the parameters. This latter uses an elegant bound for the second derivative derived in unpublished and unfinished notes of \citet{Leeuw05}. Then, to avoid overfitting issues due to the separability problem inherent of such models, we maximize a regularized maximum likelihood using the nuclear norm.

Built on recent results on the selection of the threshold  parameters for $\ell_1$ type of penalty \cite{Sardy16},  we suggest combining the \textit{universal quantile threshold} to select the rank and cross-validation to determine the amount of shrinkage.  Our results enable us to uncover some interesting insight into the balance between good selection or good prediction properties to select the threshold parameter in lasso regression \cite{Tibs:regr:1996}.


The remainder of this paper is organized as follows. After a discussion of related work,  we describe in Section \ref{sec:model} the {\em multinomial multiple correspondence analysis} model. In Section \ref{sec:majo}, we then present the minimization-majorization algorithm to estimate model's parameters. It is shown that the model and algorithm can easily be extended to allow for missing values. 
A notorious problem for this type of model is the occurrence of parameters wandering off to infinity when the estimated probabilities get close to one. This form of overfitting is avoided by adding a nuclear norm penalty.
We explain in Section \ref{sec:select} our new procedure to select the penalty parameter.

\subsection{Related Work}

The log-linear model \cite{Agresti13, Christensen90} is the golden standard to study the relationship between categorical variables. However, it  encounters difficulties with large number of variables and categories since many cells of the contingency table are equal to zero. Unsaturated log-linear models with main effects and two-way interactions could be used to restrict the number of estimated parameters, but the total number of parameters could still be substantial with many categories. 
One popular alternative consists of latent variable models that summarize the relationship between the given variables by a small number of latent ones, either categoricals or continuous.
The former case, known as latent class models \cite{Good74}, boils down to unsupervised clustering for one latent variable and nonparametric Bayesian extensions have  recently  been proposed \cite{Dunson09, Dunson12} to get rid of the difficult choice of the number of clusters. 
Our approach, MMCA, can be presented as a fixed effects latent traits models able to handle many latent variables. Other related models were studied
by \citet{deLeeuw:2006:PCA}, \citet{Li2013SEPCA}, \citet{Collins01ageneralization}, and \citet{tapio2002VEM} but dedicated to either binary data or random effects.

Finally, another popular approach to  examine the relationship between categorical variables is multiple correspondence analysis  also known as homogeneity analysis or dual scaling  \cite{Mich98, Nishisato80, leeuw14, leroux10, Green06}.  MCA can be seen as the counterpart of PCA for categorical data and involves reducing data dimensionality to provide a subspace that best represents the data in the sense of maximizing the variability of the projected points. As mentioned, it is often presented without any reference
to probabilistic models, in line with \citet{Benz73}'s idea to ``let the
data speak for itself.''  

As our model is inspired by the MCA representation, we first briefly  discuss how MCA is defined. Consider a dataset with $n$ rows and $J$ categorical variables, with $K_j$ categories each, $j=1,...,J$. The data are coded using the $n\times K$ super indicator matrix of dummy variables denoted by $\ma{G}$ with $K = \sum_j K_j$ and $g_{ijk} = 1$ if person $i$ chooses category $k$ of variable $j$ and $g_{ijk} = 0$ otherwise. A simple example of such a matrix $\ma{G}$ with $n = 10$, $J = 3$ variables with respectively $K_1 = 3, K_2 = 3,$ and $K_3 = 2$ categories is given by
    \begin{eqnarray*}
      \ma{G} = [\ma{G}_1 | \ma{G}_2 | \ma{G}_3] &=& 
      \left[{
      \begin{array}{ccc|ccc|cc}
         1 & 0 & 0 & 1 & 0 & 0 & 1 & 0 \\ 
         0 & 1 & 0 & 0 & 1 & 0 & 0 & 1 \\ 
         1 & 0 & 0 & 0 & 0 & 1 & 0 & 1 \\ 
         1 & 0 & 0 & 1 & 0 & 0 & 0 & 1 \\ 
         0 & 1 & 0 & 1 & 0 & 0 & 0 & 1 \\ 
         0 & 0 & 1 & 1 & 0 & 0 & 0 & 1 \\ 
         1 & 0 & 0 & 1 & 0 & 0 & 1 & 0 \\ 
         1 & 0 & 0 & 1 & 0 & 0 & 0 & 1 \\ 
         0 & 0 & 1 & 1 & 0 & 0 & 0 & 1 \\ 
         1 & 0 & 0 & 1 & 0 & 0 & 0 & 1
      \end{array}
      }
      \right].    
    \end{eqnarray*}

MCA, as all the principal component methods can be derived by performing  the SVD of matrices with specific row and column weights. The choice of weights ensures the property of the method such as the Chi-square interpretation of the distances between rows as well as the fact that the first principal component of MCA is the variable the most related to all the categorical variables in the sense of the $R^2$ of the analysis of variance which strengthen the presentation of MCA as an extension of PCA. 
More precisely, MCA is obtained by performing the generalized SVD \cite{Greenacre84} of the triplet \textit{data, column weights, row weights} $\left(\ma{J}\ma{G}, J^{-1}{\ma{D}_c}^{-1/2}, n^{-1}\ma{I}_n\right)$ with $\ma{D}_c$, the diagonal matrix with category frequencies and $\ma{J}_{n \times n} = (\ma{I}-n^{-1}\ma{11}')$ the row-centering matrix. It boils down to performing the following SVD: $\ma{J}\ma{G}' =  \ma{\tilde U} \boldsymbol{\Lambda}^{1/2} \ma{ \tilde V} '$ with $\ma{\tilde U}' (n^{-1}\ma{I}_n)  \ma{\tilde U} = \ma{I}$ and $ \ma{\tilde V} '  (J^{-1} {\ma{D}_c}^{-1/2}) \ma{\tilde V} = \ma{I}$.   
MCA can also be defined as  finding the best low rank approximation of $\ma{J}\ma{G}$ with a matrix of rank $p$ according to the Hilbert-Schmidt norm $\parallel {\ma{T}} \parallel^2_{\ma{D}_c^{-1/2}, \frac{1}{n}\ma{I}_n} =
\tr \left({\ma{T}}\ma{D}_c^{-1/2}{\ma{T}}^{'}{\frac{1}{n}\ma{I}_n}\right)$:
     \begin{eqnarray*}
       L_{\mathrm{MCA}}(\ma{X},\ma{A})
        &=& \|\ma{J}\ma{G} - \ma{XA'} \|^2_{\ma{D}_c^{-1/2}, \frac{1}{n}\ma{I}_n} \\
      \end{eqnarray*}
with $\ma{A}' = [\ma{A}_1'|\ldots |\ma{A}_J']$ and $\ma{A}_j$ the $K_j\times p$ matrix representing the $K_j$ categories of variable $j$. The solution is given by $\ma{A}=  \ma{\tilde V}\boldsymbol{\Lambda}^{1/4}$ and $\ma{X} =  \ma{\tilde U}\boldsymbol{\Lambda}^{1/4}$ truncated at order $p$.

Thus, the category $k$ chosen by person $i$ on variable $j$ can modeled by  
     \begin{eqnarray}
        \label{eq:mca}
       \hat g_{ijk} \approx \mu_{jk} + \ma{x}_i'\ma{a}_{jk}
     \end{eqnarray}
with  $\mu_{jk}$ the main effect for category $k$ of variable $j$, $\ma{x}_i'$ row $i$ of $\ma{X}$ and $\ma{a}_{jk}'$ row $k$ of $\ma{A}_j$. 
Equation \eqref{eq:mca} is called the \textit{reconstruction formula} in the MCA literature. 
\citet{Tenenhaus85} showed that the row sums of $\hat{\ma{G}}$ is equal to 1, which implies that the fitted values can be seen as degree of membership to the associated category. However, negative values may occur. Therefore, $\hat{g}_{ijk}$ cannot be interpreted as the probability of an individual $i$ to select category $k$ of variable $j$.

\section{Multinomial Multiple Correspondence Analysis}
\label{sec:model}

To develop a maximum likelihood approach, we consider the probability $\pi_{ijk}$ of person $i$ choosing category $k$ of variable $j$. To do so, a natural candidate is the multinomial logit or the so-called softmax function, that is,
\begin{eqnarray}
  \pi_{ijk} = \mathrm{softmax}(\bm{\theta}_i) = \frac{\exp(\theta_{ijk})}{\sum_{\ell=1}^{K_j} \exp(\theta_{ij\ell})},
  \label{for:prob}
\end{eqnarray}
where the $\theta_{ijk}$ denotes a utility that person $i$ attaches to category $k$ of variable $j$. In MMCA, the $\theta_{ijk}$ is modeled by 
\begin{eqnarray*}
  \theta_{ijk} = \mu_{jk} + \ma{x}_i'\ma{a}_{jk},
\end{eqnarray*}
very much in the same way as MCA \eqref{eq:mca} in the sense that the rank of the interaction is constrained to be $p$. 
By assuming independence between all answers of individuals on all variables and conditional independence between variables given the parameters $\theta$, the joint maximum likelihood of  MMCA is  
\begin{eqnarray*}
  \prod_{i=1}^n\prod_{j=1}^J\prod_{k=1}^{K_j} \pi_{ijk}^{g_{ijk}}.
\end{eqnarray*}

For maximum likelihood, one often minimizes the deviance by taking minus the logarithm of the probabilities  that is:
\begin{eqnarray*}
  L(\bm{\mu},\ma{X},\ma{A}) &=&-\sum_{i=1}^n\sum_{j=1}^J\sum_{k=1}^{K_j} g_{ijk}\log(\pi_{ijk})  \nonumber \\
  &=&-\sum_{i=1}^n\sum_{j=1}^J\sum_{k=1}^{K_j} g_{ijk}\log\left(\frac{\exp(\mu_{jk} + \ma{x}_i'\ma{a}_{jk})}{\sum_{\ell=1}^{K_j} \exp(\mu_{j\ell} + \ma{x}_i'\ma{a}_{j\ell})}\right) 
\end{eqnarray*}
 Without any restrictions, the parameters are not identified. Therefore, we use the following identification constraints:
\begin{itemize}
  \item $\bm{\mu}_j'\ma{1} = 0$ as adding a constant per column does not change $\pi_{ijk}$,
  \item $\ma{1}'\ma{X} = \ma{0}$ to avoid main effects estimated by $\ma{x}_i'\ma{a}_{jk}$,
  \item $\ma{X}'\ma{X} = n\ma{I}$ to take care of the rotational indeterminacy between $\ma{X}$ and the $\ma{A}_j$, and
  \item $\ma{1}'\ma{A}_j = \ma{0}'$ as adding a constant per column does not change $\pi_{ijk}$.
\end{itemize}

\begin{figure}
\definecolor{shadecolor}{rgb}{0.961, 0.961, 0.961}\color{fgcolor}

{\centering \includegraphics[width=0.4\textwidth]{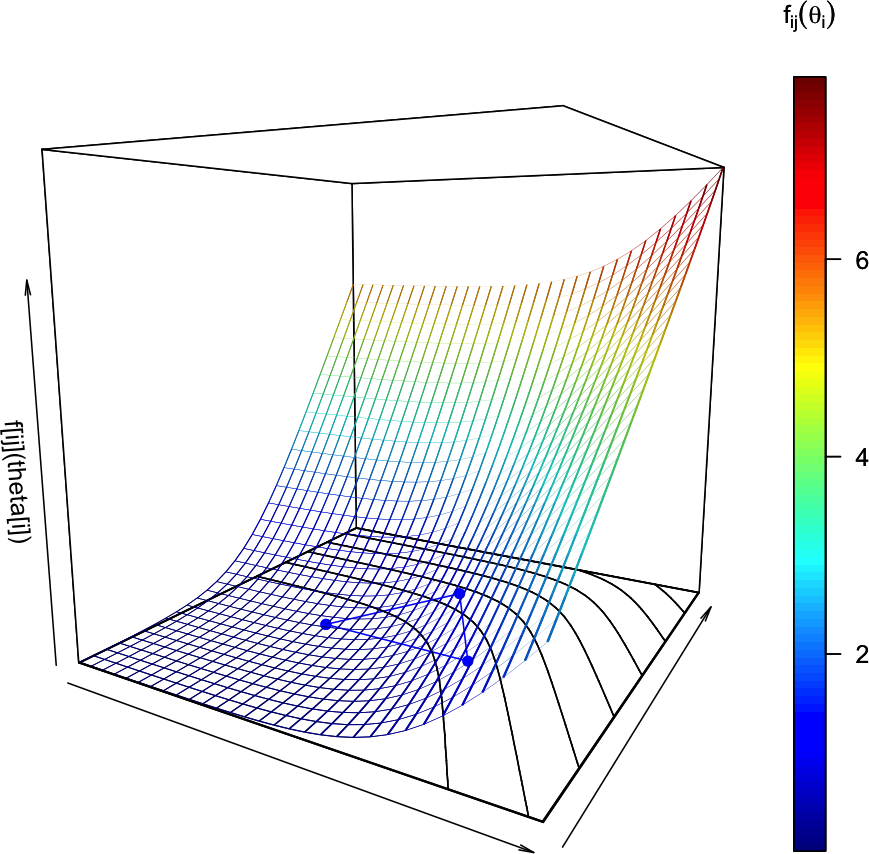} 

}

    \caption{Minus the log likelihood of one observation, $- \log(\pi_{ijk})$, for a variable with three categories, each of them shown by a vertex of the equilateral triangle, spanning a two-dimensional space.  The direction in which the left one is pointing, is the direction that asymptotically approaches a probability of one. Without additional constraints, minimization of $- \log(\pi_{ijk})$ will lead to this $\theta_{ijk}$ tending to (very) large values.}
    \label{fig:loglik}
\end{figure}

Note that the maximum dimensionality is $p^* = \min(n-1, \sum_{j=1}^J K_j - J)$. When $p = p^*$, we have a saturated model with all $\theta_{ijk} \rightarrow \infty$ for $g_{ijk} = 1$ and $\theta_{ijk} \rightarrow -\infty$ for $g_{ijk} = 0$ because all data points can be perfectly estimated. This problem does not only occur in maximum dimensionality. Even when $p \leq p^*$, we often find that several estimates for $\pi_{ijk}$ approaching one so that $\theta_{ijk}\rightarrow \infty$ when minimizing the deviance $L(\bm{\mu},\ma{X},\ma{A})$. Consider Figure~\ref{fig:loglik} that shows $-\log(\pi_{ijk})$ for individual $j$ and three categories represented by the vertices of the equilateral triangle.  The  direction of the left vertex gives an infimum of zero, that is, the further in that direction, the closer $-\log(\pi_{ijk})$ gets to zero. Therefore, there is an attraction to $\theta_{ijk}$ becoming ever larger.

This effect can be seen as a form of overfitting and a natural solution to tackle it is to add a regularization term that controls the size of the parameters in a penalized likelihood approach. 
To do so, we need to reparamatrize $\ma{XA}'$ in an SVD-type manner, that is, $\ma{XA}' = \ma{UDV}'$ with $\ma{U}$ the $n \times p$ matrix with $\ma{U}'\ma{U} = \ma{I}$ and $\ma{1'U} = \ma{0}$, $\ma{D}$ the diagonal $p \times p$ matrix with $d_{ss} \geq 0$, and $\ma{V}$ the $(\sum_j K_j) \times p$ matrix with $\ma{V}'\ma{V} = \ma{I}$, $\ma{V}_j'\ma{1} = \ma{0}$,  $\ma{V}_j' = [\ma{v}_{j1}, \ldots, \ma{v}_{jK_j}]$ represents all categories of variable $j$ and $\ma{v}_{jk}'$ is row $jk$ of $\ma{V}$  

One extensively used and studied penalty in the framework of singular values decomposition based methods is the use of the nuclear norm penalty \cite{Fazel, Srebro04, CandesSURE:2013} which is equal to the sum of the singular values $\sum_{s=1}^{p^*} d_{ss}$, leading to the penalized deviance:
\begin{eqnarray}
  L(\bm{\mu},\ma{U},\ma{D}, \ma{V})&=& -\sum_{i=1}^n\sum_{j=1}^J\sum_{k=1}^{K_j} g_{ijk}\log(\pi_{ijk}) + \lambda\left({\sum_{s=1}^{p^*} d_{ss}}\right) \label{for:def_deviance}
\end{eqnarray}
with
\begin{eqnarray*}
  \pi_{ijk} = \frac{\exp(\mu_{jk} + \ma{u}_i'\ma{D}\ma{v}_{jk})}{\sum_{\ell=1}^{K_j} \exp(\mu_{j\ell} + \ma{u}_i'\ma{D}\ma{v}_{j\ell})} 
  \label{for:def_pijk}
\end{eqnarray*}
Whenever $p = p^*$, $L(\bm{\mu},\ma{U},\ma{D}, \ma{V})$ is a convex function minimized over a convex set that has a global minimum for any choice of $\lambda > 0$.  For sufficiently large values of $\lambda$, the impact of the penalty is to set some of the smallest singular values to zero and thus often results in a lower rank solution. In addition to automatic rank selection, the nuclear-norm also shrinks the non-null singular values.

In Section \ref{sec:select}, we will provide more details on a procedure for selecting the threshold parameter $\lambda$. In the next section, a majorizing algorithm is derived for minimizing \eqref{for:def_deviance}.

\section{Majorization}
\label{sec:majo}

Majorization algorithms share as the most important property that they have a guaranteed descent, that is, in each iteration the objective function improves. Under this name, majorization was first proposed by \citet{Leeu77}. Since some time, it is probably better known under the name MM, minimization by majorization or maximization by minorization \cite{Lang04}. The principle is quite simple: in each iteration an auxiliary function (called the majorizing function) $g(\bm{\theta},\bm{\theta}_0)$ is set up that satisfies the following requirements 
\begin{enumerate}
  \item $f(\bm{\theta}_0) = g(\bm{\theta}_0,\bm{\theta}_0)$,
  \item $f(\bm{\theta}) \leq g(\bm{\theta},\bm{\theta}_0)$,
\end{enumerate}
where the current estimate $\bm{\theta}_0$ is called supporting point and $f$ is the original function to be minimized. In practice, $g(\bm{\theta},\bm{\theta}_0)$ only takes simple forms such as linear or quadratic so that its minimum $\bm{\theta}^+$ is easy to find. At its  minimum $\bm{\theta}^+$  of the majorizing function $g(\bm{\theta},\bm{\theta}_0)$, we necessarily have that $f(\bm{\theta}^+)\leq g(\bm{\theta}^+,\bm{\theta}_0)$. This leads to the so called sandwich inequality 
\begin{eqnarray*}
  f(\bm{\theta}^+)\leq g(\bm{\theta}^+,\bm{\theta}_0) \leq g(\bm{\theta}_0,\bm{\theta}_0) = f(\bm{\theta}_0)
\end{eqnarray*}
with $\bm{\theta}^+ = \mathrm{argmin~} g(\bm{\theta},\bm{\theta}_0)$ proving that an update of the majorizing function also improves $f$ until no improvement is possible. 

The advantage of majorization is that in contrast to line search methods such as steepest descent or (quasi-)Newton methods, there is no need for a possibly computationally expensive steplength procedure to guarantee descent.  

To derive the majorizing algorithm, the following steps are taken. First, a quadratic majorizing function is derived for the elements of the deviance denoted $f_{ij}(\thetai)$. Then, a majorizing function is given for the penalized deviance $L(\bm{\mu},\ma{U},\ma{D},\ma{V})$ in (\ref{for:def_deviance}). 
To choose $\lambda$, we wish to use cross-validation and thus we need to be able to minimize the penalized deviance in the presence of  missing values. This requires an additional majorization function describe in a third step. 
The last step consists for each of the four sets of parameters in  deriving an update for the parameters. Finally, an overview of the entire algorithm is presented. \\

\textit{First step: main majorizing function.} \\
The first step is finding a majorizing function for a single term of the deviance function 
\begin{eqnarray}
\label{eq:majofij}
     f_{ij}(\thetai) = -\sum_{k=1}^{K_j}g_{ijk}\log(\pi_{ijk}) = -\sum_{k=1}^{K_j}g_{ijk}\log\left(\frac{\exp(\theta_{ijk})}{\sum_{\ell=1}^{K_j} \exp(\theta_{ij\ell})}\right).
\end{eqnarray}
To do so, an explicit expression of its first derivative is needed
\begin{eqnarray}
\label{eq:grad}
     \nabla f_{ij}(\thetai) = \ma{g}_{ij} - \bm{\pi}_{ij},
\end{eqnarray}
where $\ma{g}_{ij}'$ and $\bm{\pi}_{ij}'$ represent row $i$ of $\ma{G}_j$ and $\bm{\Pi}_j$, respectively with $\bm{\Pi}$ the $n \times \sum_{j=1}^J K_j$ matrix of probabilities $\bm{\Pi}$ having elements $\pi_{ijk}$. Now, the following theorem coming from unfinished notes of \citet{Leeuw05} presents a quadratic majorizing function of \eqref{eq:majofij}.

\begin{theorem}
\label{theo:majo}
\begin{eqnarray*}
        g_{ij}(\thetai,\bm{\theta}_i^{(0)}) = f_{ij}(\bm{\theta}_i^{(0)}) + (\thetai - \bm{\theta}_i^{(0)})'\nabla f_{ij}(\bm{\theta}_i^{(0)}) + 1/4 \|\thetai - \bm{\theta}_i^{(0)}\|^2
\end{eqnarray*}
is a majorizing function of $f_{ij}(\thetai)$.
\end{theorem}

\begin{figure}
\definecolor{shadecolor}{rgb}{0.961, 0.961, 0.961}\color{fgcolor}

{\centering \includegraphics[width=0.4\textwidth]{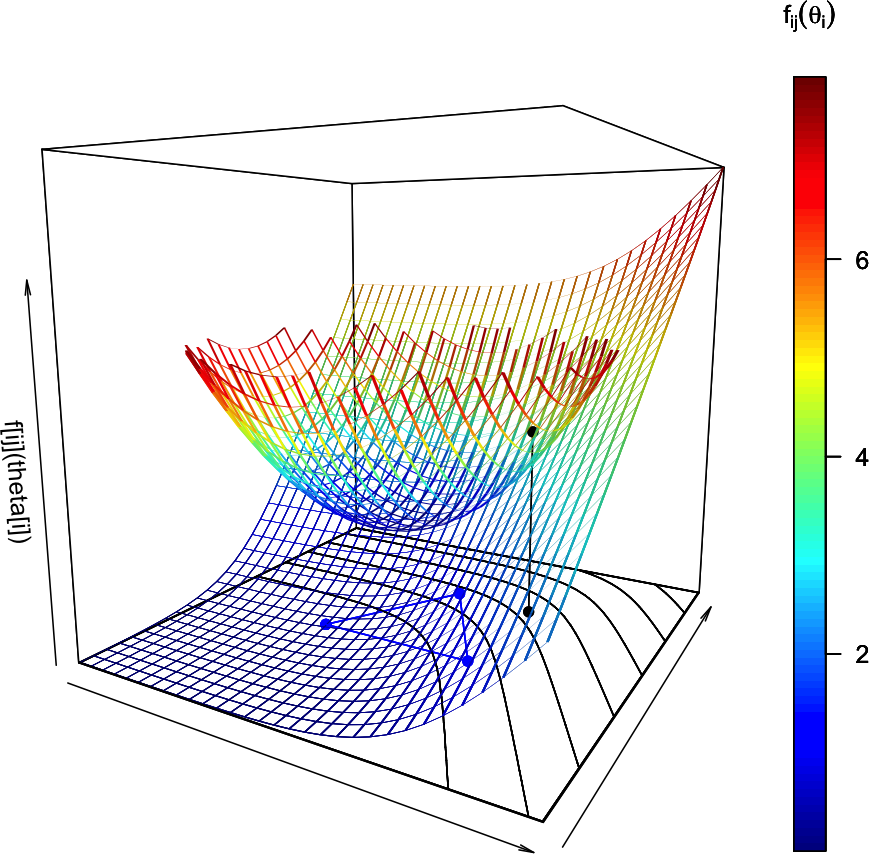} 

}

  \caption{Quadratic majorizing function $g_{ij}(\thetai,\bm{\theta}_i^{(0)})$ that touches $f_{ij}(\thetai)$ at $f_{ij}(\thetai)$ at $\bm{\theta}_i^{(0)}$ (black vertical line) and is located above $f_{ij}(\thetai)$ elsewhere.}
  \label{fig:maj}
\end{figure}

In the appendix, this thereom is proved in a slightly different way compared to \citet{Leeuw05}.
Figure~\ref{fig:maj} gives an example of the quadratic majorizing function. 

~\\
\textit{Second step: combining majorization.} \\
To obtain a majorizing function for the deviance $L(\bm{\mu},\ma{U},\ma{D},\ma{V})$, one needs to sum $f_{ij}(\thetai)$ and $g_{ij}(\thetai,\bm{\theta}_i^{(0)})$ over all $ijk$ and to replace the gradient by its value \eqref{eq:grad}, it gives the majorizing function
\begin{eqnarray}
    L(\bm{\mu},\ma{U},\ma{D},\ma{V})  \leq\frac{1}{4}\sum_{i=1}^n\sum_{j=1}^J\sum_{k=1}^{K_j} g_{ijk}(z_{ijk}-\theta_{ijk})^2 
    + \lambda\left({\sum_{s=1}^pd_{ss}}\right) + c 
    \label{for:dev_maj_first_sum}
\end{eqnarray}
with
\begin{eqnarray*}
  z_{ijk} = \mu_{jk}^{(0)} + {\ma{u}_i^{(0)}}'\ma{D}^{(0)}{\ma{v}_{jk}^{(0)}}  + 2(g_{ijk} - \pi_{ijk}) 
\end{eqnarray*}
and $c$ containing parameters that do not depend on $\bm{\mu},\ma{U},\ma{D},$ and $\ma{V}$. \\

\textit{Third step: including missing values.}\\
Here, we assume that missing values occur due to the most simple process, that is, missing completely at random. 
If a person $i$ has a missing value on variable $j$, we can consider that $g_{ijk} = 0$ for $k=1$ to $K_j$. In this way, the missing value does not contribute to the penalized deviance (\ref{for:def_deviance}).
When not all $\ma{g}_{ij}$ have a one, the majorizing function is a weighted least-squares function. With rank restrictions (as we have when $p \neq p^*$), such a weighted least-squares problem is more difficult to solve. One additional majorizing step is needed. Let $\ma{W}$ be an $n \times (\sum_{j=1}^J K_j)$ matrix that has $\ma{w}_{ij}' = \ma{1}'$ if person $i$ does not have a missing value on variable $j$ and $\ma{w}_{ij}' = \ma{0}'$ otherwise. Using results from \citet{Kiers97}, we have that
\begin{eqnarray}
  (w_{ijk}-1)(\theta_{ijk} - \theta_{ijk}^{(0)})^2&\leq&0 \nonumber \\
  w_{ijk}\theta_{ijk}^2&\leq&\theta_{ijk}^2 
  - 2\theta_{ijk}(1-w_{ijk})\theta_{ijk}^{(0)}
  + (1-w_{ijk})\left({\theta_{ijk}^{(0)}}\right)^2.
  \label{for:missing}
\end{eqnarray}
Combining the majorization in \eqref{for:dev_maj_first_sum} and (\ref{for:missing}) gives
\begin{eqnarray}
    L(\bm{\mu},\ma{U},\ma{D},\ma{V})   \leq   \frac{1}{4}\|\ma{Z} -(\ma{1}\bm{\mu}' + \ma{UDV}')\|^2  + \lambda\left({\sum_{s=1}^pd_{ss}}\right) + c \nonumber\\
    \leq  \frac{1}{4}\|\ma{Z} -\ma{1}\bm{\mu}'\|^2  +\frac{1}{4}\|\ma{JZ} -\ma{UDV}'\|^2  
    + \lambda\left({\sum_{s=1}^pd_{ss}}\right) + c 
    \label{for:maj_dev}
\end{eqnarray}
with
\begin{eqnarray*}
  \ma{Z} = [(\ma{1}{\bm{\mu}^{(0)}}' + \ma{U}^{(0)}\ma{D}^{(0)}{\ma{V}^{(0)}}')  + 2(\ma{G} - \ma{W}\odot\bm{\Pi})]\ma{J}, 
\end{eqnarray*}
with $c = L(\bm{\mu}^{(0)},\ma{U}^{(0)},\ma{D}^{(0)},\ma{V}^{(0)}) - 1/4 \|\ma{Z}\|^2$ and $\odot$ means the elementwise multiplication of two matrices. \\ 

\textit{Fourth step:  update for the four sets of parameters.}\\ 
The update for $\bm{\mu}$ simply amounts to minimizing $\|\ma{Z} -\ma{1}\bm{\mu}'\|^2$ which is done by 
\begin{eqnarray*}
  \bm{\mu} = n^{-1}\ma{Z}'\ma{1}.
  \label{for:update_mu}
\end{eqnarray*}

To update $\ma{U}$ and $\ma{V}$ it is sufficient to  minimize the crosspruduct term $-\tr \ma{Z}'\ma{J} \ma{UDV}'$ because the quadratic term in (\ref{for:maj_dev}) disappears due to their orthonormality restrictions. Let $\ma{JZ} = \ma{P}\bm{\Phi}\ma{Q}'$ be the SVD. So-called Kristof lower bounds are available for linear sums of orthonormal matrices, that is,
\begin{eqnarray*}
  -\tr \ma{Z'J} \ma{U} \ma{D} \ma{V}' &=&
           -\tr  \ma{Q}\bm{\Phi}\ma{P}'\ma{U}\ma{D}\ma{V}'
           = -\tr \bm{\Phi}(\ma{P}' \ma{U})   \ma{D} (\ma{V}'\ma{Q}) \\
           &\geq& -\tr \bm{\Phi}(\ma{I})   \ma{D} (\ma{I})
\end{eqnarray*}
and the lower bound is attained at 
\begin{eqnarray*}
  \ma{U} = \ma{P} \mbox{~and~} \ma{V}=\ma{Q}.
  \label{for:update_UV}
\end{eqnarray*}

To update $\ma{D}$, we write the relevant part of the majorizing function (\ref{for:maj_dev}) as
\begin{eqnarray}
  \sum_{s=1}^p \left[{(\phi_{ss} - d_{ss})^2 + \lambda\left({d_{ss} }\right) }\right]
  \label{for:maj_dev_D}
\end{eqnarray}
subject to $d_{ss} \geq 0$. It can be verified that the update 
\begin{eqnarray}
  d_{ss} = \max(0, \phi_{ss} - \lambda)
  \label{for:update_D}
\end{eqnarray}
is optimal for to minimize (\ref{for:maj_dev_D}).

A summary of the updates and the majorization algorithm for MMCA is given by Algorithm~\ref{alg:maj}. After convergence, one may choose to set $\ma{X} = n^{1/2}\ma{UD}^{1/4}$ and $\ma{A} = n^{-1/2}\ma{VD}^{1/4}$.

\begin{algorithm}[t]
  \SetAlgoLined
  \KwData{$\ma{G}, p, \lambda, \alpha, \epsilon$} 
  \KwResult{$\bm{\mu}, \ma{U}, \ma{D}, \ma{V}$}
  $t = 0$\;
  Compute $\ma{W}$ from missing values in $\ma{G}$\;
  Initialize $\bm{\mu}$: $\bm{\mu}^{(0)} = n^{-1}\ma{J}_c\ma{G}'\ma{1}$ \;
  Compute the SVD of $\ma{JGJ}_c$: $\ma{JGJ}_c = \ma{P}\bm{\Phi}\ma{Q}'$\;
  Initialize $\ma{U}$: $\ma{U}^{(0)} = \ma{P}$ \;
  Initialize $\ma{V}$: $\ma{V}^{(0)} = \ma{Q}$ \;
  Initialize $\ma{D}$: $d_{ss}^{(0)} = \max(0, \phi_{ss} - \lambda)$ \;
  Compute $\bm{\Pi}$ by (\ref{for:def_pijk}) \;
  Compute $L^{(0)} = L(\bm{\mu}^{(0)},\ma{U}^{(0)},\ma{D}^{(0)}, \ma{V}^{(0)})$ \;
  \While{$t=0$ or $(L^{(t)} - L^{(t-1)})/L^{(t)} \geq \epsilon$}{
    $t = t + 1$\;
    $\ma{Z} = [(\ma{1}{\bm{\mu}^{(t-1)}}' + \ma{U}^{(t-1)}\ma{D}^{(t-1)}{\ma{V}^{(t-1)}}')  + 2(\ma{G} - \ma{W}\odot\bm{\Pi})]\ma{J}_c$ \;
    Compute update $\bm{\mu}$: $\bm{\mu}^{(t)} = n^{-1}\ma{Z}'\ma{1}$ \;
    Compute the SVD of $\ma{JZ}$: $\ma{JZ} = \ma{P}\bm{\Phi}\ma{Q}'$\;
    Update $\ma{U}$: $\ma{U}^{(t)} = \ma{P}$ \;
    Update $\ma{V}$: $\ma{V}^{(t)} = \ma{Q}$ \;
    Update $\ma{D}$: $d_{ss} = (1 + \max(0, \phi_{ss} - \lambda)$ \;
    Compute $\bm{\Pi}$ by (\ref{for:def_pijk}) \;
    Compute $L^{(t)} = L(\bm{\mu}^{(t)},\ma{U}^{(t)},\ma{D}^{(t)},\ma{V}^{(t)})$ \;
  }
\caption{The majorizing algorithm for MMCA. $\epsilon$ is here a small positive value, for example, $\epsilon = 10^{-8}$.}
\label{alg:maj}
\end{algorithm}

\section{Properties and Interpretation}
\label{sec:properties}

At a stationary point of the algorithm (in practical cases a local minimum) the solution satisfies several properties. 

\begin{property}
The main effect $\mu_{jk}$ can be interpreted as the log-odds against zero (that is, all categories are equally likely).
\label{prop:mu}
\end{property}

\begin{proof}
Assuming that all other parameters are equal zero, then the probability of person $i$ choosing category $k$ of variable $j$ equals $\pi_{ijk} = \exp(\mu_{jk})/\sum_{\ell=1}^{K_j} \exp(\mu_{j\ell})$. The probability for $\mu_{jk} = 0$ equals $\pi_{ijk}^* = \exp(0)/\sum_{\ell=1}^{K_j} \exp(\mu_{j\ell})$. The odds are
\begin{eqnarray*} 
  \frac{\pi_{ijk}}{\pi_{ijk}^*} = \frac{\left(\frac{\exp(\mu_{jk})}{\sum_{\ell=1}^{K_j} \exp(\mu_{j\ell})}\right)}{\left(\frac{\exp(0)}{\sum_{\ell=1}^{K_j} \exp(\mu_{j\ell})}\right)} = \exp(\mu_{jk})
\end{eqnarray*}
and the log odds equals
\begin{eqnarray*}
  \log\frac{\pi_{ijk}}{\pi_{ijk}^*} = \mu_{jk}.
\end{eqnarray*}
\end{proof}

Let $\ma{X} = n^{1/2}\ma{U}$ and $\ma{A} = n^{-1/2} \ma{VD}$. In a bilinear biplot, the projection interpretation of $\ma{x}_i'\ma{a}_{jk}$ implies that a point $\ma{x}_i$ representing individual $i$ should be projected onto the vector $\ma{a}_{jk}$ representing category $k$ of variable $j$ and multiplied by the length $\|\ma{a}_{jk}\|$ of $\ma{a}_{jk}$. 

\begin{property}
The interaction effect $\ma{x}_i'\ma{a}_{jk}$ can be interpreted as the log-odds against zero that would be obtained by projecting a person  onto $\ma{a}_{jk}$ at the origin.
\end{property}

\begin{proof}
Assume all parameters equal to zero except $\ma{x}_i'\ma{a}_{jk}$.  Following the same reasoning as in the proof of Property~\ref{prop:mu} gives the desired result.
\end{proof}

$L$ can be rewritten as squared Euclidean distance between $\ma{x}_i$ and $\ma{a}_{jk}$.

\begin{property}
The fitted category probabilities sum to the observed frequency of that category, that is, $\ma{1}'\bm{\Pi} = \ma{1}'\ma{G}$.
\end{property} 

\begin{proof}
At convergence we must have that $\bm{\mu} = \bm{\mu}^{(0)}$. Therefore, 
\begin{eqnarray*}
  \bm{\mu} &=& n^{-1}\ma{Z}'\ma{1} \\
    &=& n^{-1}\left({\bm{\mu}\ma{1}' +  2(\ma{G} - \bm{\Pi})'}\right)\ma{1} \\
    &=& \bm{\mu} + 2n^{-1}(\ma{G} - \bm{\Pi})'\ma{1}
\end{eqnarray*} 
which can only hold if $(\ma{G} - \bm{\Pi})'\ma{1} = \ma{0}$, or, equivalently, if $\ma{G}'\ma{1} = \bm{\Pi}'\ma{1}$. This completes the proof.
\end{proof}

The  weighted centroids of $\ma{X}$ with the weights being the probability of choosing the category is given by
\begin{eqnarray*}
   \mbox{Diag}(\ma{1}'\ma{G})^{-1}\bm{\Pi}'\ma{X}
\end{eqnarray*} 
where $\mbox{Diag}(\ma{1}'\ma{G})$ is the diagonal matrix of observed counts for each of the categories. 

\begin{property}
$\ma{B} = K\mbox{Diag}(\ma{1}'\ma{G})^{-1}\ma{A}$ is a measure for the bias of the category centroids, that is, the difference between the weighted and unweighted centroids, that is,
\begin{eqnarray*}
  \ma{B} = K\mbox{Diag}(\ma{1}'\ma{G})^{-1}\ma{A} = \mbox{Diag}(\ma{1}'\ma{G})^{-1}(\ma{G}-\bm{\Pi})'\ma{X}.
\end{eqnarray*} 
\end{property}

\begin{proof}
After convergence, it must also be true that the update of a single set of parameters yields the same solution. Consider the update of $\ma{A}$ for a given $\ma{X}$. The majorizing algorithm says that at convergence we have ... 
\end{proof}


\begin{property}
The three parameter logistic model and its multidimensional variant are special cases of MMCA.
\end{property}

\begin{figure}
\definecolor{shadecolor}{rgb}{0.961, 0.961, 0.961}\color{fgcolor}

{\centering \includegraphics[width=0.4\textwidth,height=3cm]{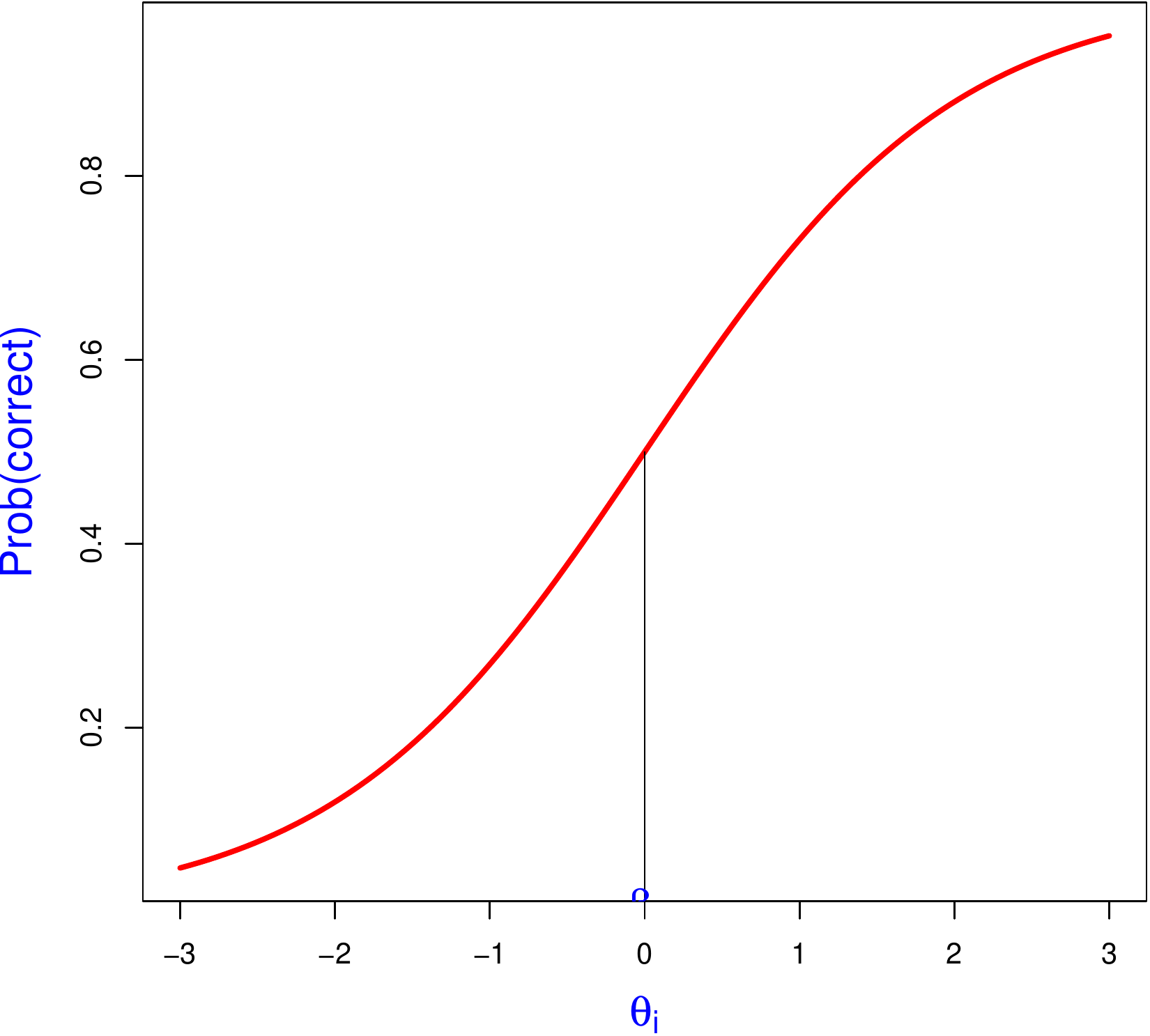} 

}

\caption{Example of the item characteristic curve used by the three parameter logistic IRT model.}
\label{fig:IRT}
\end{figure}

\begin{proof}
In the three parameter logistic IRT model, the probability that individual $i$ gives a correct answer to item $j$ is equal to $(1+e^{-\gamma_{ij}})^{-1}$ with $\gamma_{ij} = \alpha_j\theta_i-\beta_j$ and $\theta_i$ is the ability of individual $i$, $\beta_j$ the item difficulty parameter of item $j$, and $\alpha_j$ the item discrimination parameter of item $j$. Figure~\ref{fig:IRT} gives the item characteristic curve by the three parameter logistic model.

The multidimensional IRT model can be written as
\begin{eqnarray*}
  \gamma_{ij} &=& -\beta_j + \sum_{s=1}^p \theta_{is}\alpha_{js} = -\beta_j + \bm{\theta}_{i}'\bm{\alpha}_{j} \\
    &=& \mu_j +  \bm{x}_{i}'\bm{a}_{j}.
\end{eqnarray*}
If all variables are binary (all $K_j=2$) and $\lambda = 0$, then the joint maximum likelihood approach to the multidimensional IRT model is equivalent to MMCA. 
\end{proof}

The overfitting problem in multidimensional IRT can be easily tackled by the penalty term of the MMCA approach. 

\section{Selecting the Regularization Parameters}
\label{sec:select}

Selecting the threshold parameter $\lambda$ is crucial in the method. 
We suggest a two steps procedure. First, we present an appropriate way to estimate the rank $p$ of the interaction. Then, we select $\lambda$ using cross-validation for a specified rank.  To understand the rationale of such an approach, let's start by reviewing some key concepts and recent results to select the threshold parameter in the framework of $\ell_1$ based penalty.

\citet{Sardy16} highlighted that lasso regression is very often used for its screening properties \cite{Buhl11} and its ability to select variables since it thresholds the coefficients estimate towards 0. 
However, current methods for selecting $\lambda$ such as cross-validation or Stein unbiased risk estimation \cite{Zou07, Tib12} focus on good predictive performances. They pointed out that the optimal threshold for prediction is typically different from the optimal one for screening and somewhat smaller which generally leads to too complex models. 
Thus, \citet{Sardy16} suggested a \textit{quantile universal threshold} (QUT) that guarantee variable screening with high probability.  
However, their threshold is not appropriate for prediction since it biases too much the estimates.  
The rationale of the QUT approach is to selecting the threshold at the bulk edge of what a threshold should be under the null model that all the variables coefficients are equal to zero.  \citet{Josse15} used a similar idea of a null model in the context of low rank matrix estimation to estimate the support, \textit{i.e.} the rank.  

We can built on both works to suggest a strategy to select $\lambda$ aiming at good rank recovery.
The estimated rank is determined with the threshold $\lambda$ since any empirical singular value $d_{ss}$ smaller than $\lambda$ is set to zero by (\ref{for:update_D}).
The procedure work as follows: for a given data set $\ma{G}$, we estimate the main effect $\bm{\mu}$ and generate data under the MMCA model \eqref{for:prob} and the null hypothesis of no interaction $\ma{X}\ma{A'}=0$; then we apply the whole procedure (Algorithm ~\ref{alg:maj}) and return the first singular value $d_{1,1}$. This procedure is repeated 10000 times. 
Then, we  use the $(1-\alpha)$-quantile of the distribution of the largest empirical singular value under the null hypothesis to determine the selected threshold.
Following  \citet{Dono94b} and \citet{Sardy16} who implicitly used level of order $\alpha=O(1/\sqrt{\log n})$ when $n=J$,
we choose a similar level tending to zero with the maximum of $n$ and $(K-J)$.
This leads to the definition of the \textit{quantile universal threshold} for estimating the rank in MMCA:
\begin{equation} \label{eq:univ}
\lambda_{\max(n, K-J)}=\sigma F_{\Lambda_1}^{-1}\left(1-\frac{1}{\sqrt{\log(\max(n, K-J))}}\right),
\end{equation}
where $F_{\Lambda_1}$ is the cumulative distribution function of the largest singular value. 

Although this $\lambda$ enjoy very good property of rank recovery, its value is far too large and it shrinks too much the sequel singular values. 
That's why, we only  use this procedure to select the rank and then for a given rank $p$, we select $\lambda$ to minimize the penalized deviance \ref{for:def_deviance} using cross-validation. 
Note that it is in agreement with the common practice in lasso, where often lasso is used to select variables and then ordinary least squares are applied. Here we claim that in our setting we still need to shrink after the selection step.

Leave-one-out cross-validation, first consists in removing one cell of the categorical data matrix for one individual $i$ on a variable $j$ leading to a row $\ma{g}_{ij}$ with missing values. Then, it consists in predicting its value using the estimator obtained from the dataset that excludes these. The value of the predicted elements is denoted $\ma{\pi} _{ij}^{-ij}$. Finally, the deviance is computed with these predicted values. The operation is repeated for all the cells in the categorical data and for a grid for $\lambda$.  The  value of  $\lambda$ that minimizes the deviance is selected. Such method is of course computationally intensive and so we have implemented a  $K$-fold strategy and a parallelized version using the different cores of the machine.

\section{Conclusion and Discussion}

We introduced the \textit{multinomial multiple correspondence analysis} to model the dependence between categorical variables using a low-rank representation of the data.
The challenges in the estimation of  the parameters were adressed with a majorization algorithm combined with a nuclear norm penalization. The universal threshold allows to accurately estimate the rank while cross-validation ensures good selection of the thresholding parameter. There appear to be many potential applications for our methods since it is possible to scale majorization algorithm to the difficult cases of large sparse data sets. 
One drawback is that contrary to MCA, the biplot representation does not enjoy the centroid property. It means that categories are not at the barycenter of the individuals which have selected the categories. Thus, biplot may be less easy to interpret. Note that in terms of graphical outputs MCA enjoys many nice interpretation but as already mentionned, the estimated values for the probabilities sum to one but can take negative values and thus does not represent a proper way to model the probabilities of individuals taken categories.

We finish by discussing some opportunities for further research. 
We used a two-step approach to select the regularization parameter, after the hard-thresholding step we still shrink with a soft-thresholding approach. This is in the same vein than selecting variables with LASSO and using ordinary least-squares except that we use in the second step an additional regularization. This may indicate that we should consider other penalties and scheme of regularization allowing compromise between hard and soft thresholding. In the framework of low rank matrix approximation with Gaussian noise, recent works showed that the signal was better recovered when non-linear transformation of the singular values were applied.  For instance \citet{Shabalin2013} and \citet{OS:2014} gave an explicit transformation in a asymptotic framework where both $n$ and $J$ tend to infinity while \citet{Josse15} considered a finite sample situation and suggested an adaptive penalty inspired by adaptive LASSO. Their method provides a large family of thresholding function that goes between hard and soft thresholding. Extending these ideas for categorical data provide some challenges both to get results outside the Gaussian case or to include easily different penalties in the majorization algorithm.

The cross-validation procedure and the capability of the method to handle missing values encourage investigating the use of this method to handle missing values in a broader framework. Indeed, one main strategy avalaible to deal with missing values \cite{Little02} consists in using imputation methods, it means replacing the missing values by plausible values to get a completed data on which any statistical analysis can be applied.  More precisely, the recomemend strategy is to use multiple imputation \citet{Rubin87} where multiple values are predicted for each missing entrie to take into account the uncertainty of prediction in the sequel analyses.  Many multiple imputation techniques are available for continuous data \cite{VB12} but the litterature is less abundant for categorical ones. It can be explained by the difficulty to get an imputation model and an estimation strategy which can handle large number of categories per variable, a large number of variables or a small number of individuals. That's why we may expect some interesting results in this direction.

\appendix
\section{Proofs}

\begin{proof}[Theorem \ref{theo:majo}]
To prove that $g_{ij}(\thetai,\bm{\theta}_i^{(0)}) = = f_{ij}(\bm{\theta}_i^{(0)}) + (\thetai - \bm{\theta}_i^{(0)})'\nabla f_{ij}(\bm{\theta}_i^{(0)}) + 1/4 \|\thetai - \bm{\theta}_i^{(0)}\|^2$ is a majorizing function of $f_{ij}(\thetai)$ two conditions  must hold. The first one is that $f_{ij}(\thetai) = g_{ij}(\thetai,\bm{\theta}_i^{(0)})$ at the supporting point $\thetai = \bm{\theta}_i^{(0)}$. Working out  $g_{ij}(\bm{\theta}_i^{(0)},\bm{\theta}_i^{(0)})$ trivially shows that it is equal to $f_{ij}(\bm{\theta}_i^{(0)})$ thereby confirming the first assumption.

The second requirement is that $f_{ij}(\thetai) \leq g_{ij}(\thetai,\bm{\theta}_i^{(0)})$ or, equivalently, $h_{ij}(\thetai)  =  g_{ij}(\thetai,\bm{\theta}_i^{(0)}) -  f_{ij}(\thetai) \geq 0$ for all $\thetai$ with equality for $\thetai=\bm{\theta}_i^{(0)}$. Equality was already proven above. The inequality is automatic if (i) the gradient of $h_{ij}(\thetai)$ is zero at the supporting point $\bm{\theta}_i^{(0)}$ and (ii) $h_{ij}(\thetai)$ is convex. The gradient of $h_{ij}(\thetai)$ is given by 
\begin{eqnarray*}
     \nabla h_{ij}(\thetai) &=& \nabla g_{ij}(\thetai,\bm{\theta}_i^{(0)}) - \nabla h_{ij}(\thetai) \\
     & = & \nabla f_{ij}(\bm{\theta}_i^{(0)}) + 1/2(\thetai - \bm{\theta}_i^{(0)}) - \nabla f_{ij}(\thetai)
\end{eqnarray*}
so that
\begin{eqnarray*}
     \nabla h_{ij}(\bm{\theta}_i^{(0)}) &=&  \nabla f_{ij}(\bm{\theta}_i^{(0)}) + 1/2(\bm{\theta}_i^{(0)} - \bm{\theta}_i^{(0)}) - \nabla f_{ij}(\bm{\theta}_i^{(0)}) = \ma{0}.
\end{eqnarray*}
and Condition (i) is satisfied. The Hessian of $f_{ij}(\thetai)$ is given by 
\begin{eqnarray*}
     \nabla^2 f_{ij}(\thetai) &=&  \textrm{Diag}(\bm{\pi}_{ij}) -\bm{\pi}_{ij}\bm{\pi}_{ij}'
\end{eqnarray*}
and that of 
\begin{eqnarray*}
     \nabla^2 g_{ij}(\thetai,\bm{\theta}_i^{(0)}) &=& 1/2\ma{I}
\end{eqnarray*}
so that the Hessian of $h_{ij}(\thetai)$
\begin{eqnarray*}
     \nabla^2 h_{ij}(\thetai) &=& \nabla^2 g_{ij}(\thetai,\bm{\theta}_i^{(0)}) - \nabla^2 f_{ij}(\thetai) =  1/2\ma{I} - (\textrm{Diag}(\bm{\pi}_{ij}) -\bm{\pi}_{ij}\bm{\pi}_{ij}').
\end{eqnarray*}
For Condition (ii), convexity of $h_{ij}(\thetai)$, to hold it suffices to prove that $\nabla^2 h_{ij}(\thetai)$ is positive semidefinite for all $\thetai$, or, equivalently, that all eigenvalues of $\textrm{Diag}(\bm{\pi}_{ij}) -\bm{\pi}_{ij}\bm{\pi}_{ij}'$ are smaller that $1/2$. An upper bound of the eigenvalues can be obtained by Gerschgorin disks which say that the eigenvalue $\phi$ is always smaller than a diagonal element plus the sum of its absolute off-diagonal row (or column) values, i.e., 
\begin{eqnarray}
  \phi 
  &\leq& \pi_{ijk} - \pi_{ijk}^2 + \pi_{ijk}\sum_{\ell \neq k} \pi_{ij\ell} \\
  &=& \pi_{ijk} - \pi_{ijk}^2 + \pi_{ijk}\sum_{\ell=1}^{K_j} \pi_{ij\ell} - \pi_{ijk}^2 \\
  &=&2(\pi_{ijk} - \pi_{ijk}^2) = 2\pi_{ijk}(1 - \pi_{ijk}).
\end{eqnarray}
It can be verified that $2\pi_{ijk}(1 - \pi_{ijk})$ reaches its maximum of $1/2$ at $\pi_{ijk} = 1/2$ so that the maximum eigenvalue  of $\nabla^2 f_{ij}(\thetai)$ is always smaller than (or equal to) $\phi = 1/2$ and, thus, $\nabla^2 h_{ij}(\thetai)$ is positive semidefinite and $h_{ij}(\thetai)$ is convex. 
\end{proof}

\bibliographystyle{plainnat}  

\bibliography{josse} 

\begin{thebibliography}{36}
\providecommand{\natexlab}[1]{#1}
\providecommand{\url}[1]{\texttt{#1}}
\expandafter\ifx\csname urlstyle\endcsname\relax
  \providecommand{\doi}[1]{doi: #1}\else
  \providecommand{\doi}{doi: \begingroup \urlstyle{rm}\Url}\fi

\bibitem[Agresti(2013)]{Agresti13}
A.~Agresti.
\newblock \emph{Categorical Data Analysis, 3rd Edition}.
\newblock Wiley, 2013.

\bibitem[Benz\'ecri(1973)]{Benz73}
J.~P. Benz\'ecri.
\newblock \emph{L'analyse des donn\'ees. {Tome II}: {L}'analyse des
  correspondances}.
\newblock Dunod, 1973.

\bibitem[Bhattacharya and Dunson(2012)]{Dunson12}
A.~Bhattacharya and D.~B. Dunson.
\newblock Simplex factor model for multivariate unordered categorical data.
\newblock \emph{Journal of the American Statistical Association}, 107
  (497):\penalty0 362--377, 2012.

\bibitem[B\"{u}hlmann and van~de Geer(2011)]{Buhl11}
P.~B\"{u}hlmann and S.~van~de Geer.
\newblock \emph{Statistics for High-Dimensional Data}.
\newblock Springer-Verlag, 2011.

\bibitem[Buntine(2002)]{tapio2002VEM}
Wray Buntine.
\newblock Variational extensions to {EM} and multinomial {PCA}.
\newblock In Tapio Elomaa, Heikki Mannila, and Hannu Toivonen, editors,
  \emph{Machine Learning: {ECML} 2002}, volume 2430 of \emph{Lecture Notes in
  Computer Science}, pages 23--34. Springer Berlin Heidelberg, 2002.

\bibitem[Candes et~al.(2013)Candes, Sing-Long, and Trzasko]{CandesSURE:2013}
E.~J. Candes, C.~A. Sing-Long, and J.~D. Trzasko.
\newblock Unbiased risk estimates for singular value thresholding and spectral
  estimators.
\newblock \emph{IEEE Transactions on Signal Processing}, 61\penalty0
  (19):\penalty0 4643--4657, 2013.

\bibitem[Christensen(2010)]{Christensen90}
R.~Christensen.
\newblock \emph{Log-Linear Models}.
\newblock Springer-Verlag, New York, 2010.

\bibitem[Collins et~al.(2001)Collins, Dasgupta, and
  Schapire]{Collins01ageneralization}
Michael Collins, Sanjoy Dasgupta, and Robert~E. Schapire.
\newblock A generalization of principal component analysis to the exponential
  family.
\newblock In \emph{Advances in Neural Information Processing Systems}. MIT
  Press, 2001.

\bibitem[De~Leeuw(2005)]{Leeuw05}
J.~De~Leeuw.
\newblock Gifi goes logistic.
\newblock Technical report, Department of Statistics, University of California,
  Los Angeles, 2005.
\newblock URL \url{http://gifi.stat.ucla.edu/janspubs}.

\bibitem[De~Leeuw(2014)]{leeuw14}
J.~De~Leeuw.
\newblock History of non linear principal component analysis.
\newblock In J.~Blasius and M.~J. Greenacre, editors, \emph{Visualization and
  Verbalization of Data}. Chapman \& Hall, 2014.

\bibitem[De~Leeuw and Heiser(1977)]{Leeu77}
J.~De~Leeuw and W.~J. Heiser.
\newblock Convergence of correction matrix algorithms for multidimensional
  scaling.
\newblock In J.~C. Lingoes, E.~Roskam, and I.~Borg, editors, \emph{Geometric
  Representations of Relational Data}, pages 735--752. Mathesis Press, 1977.

\bibitem[De~Leeuw(2006)]{deLeeuw:2006:PCA}
Jan De~Leeuw.
\newblock Principal component analysis of binary data by iterated singular
  value decomposition.
\newblock \emph{Computational Statistics and Data Analysis}, 50\penalty0
  (1):\penalty0 21--39, 2006.

\bibitem[Donoho and Johnstone(1994)]{Dono94b}
D.~L. Donoho and I.~M. Johnstone.
\newblock Ideal spatial adaptation via wavelet shrinkage.
\newblock \emph{Biometrika}, 81:\penalty0 425--455, 1994.

\bibitem[Dunson and Xing(2009)]{Dunson09}
D.~B. Dunson and C.~Xing.
\newblock Nonparametric {Bayes} modeling of multivariate categorical data.
\newblock \emph{Journal of the American Statistical Association}, 104:\penalty0
  1042--1051, 2009.

\bibitem[Fazel(2002)]{Fazel}
M.~Fazel.
\newblock \emph{Matrix Rank Minimization with Applications}.
\newblock PhD thesis, Stanford University, 2002.

\bibitem[Gavish and Donoho(2014)]{OS:2014}
M.~Gavish and D.~L. Donoho.
\newblock Optimal shrinkage of singular values.
\newblock \emph{arXiv:1405.7511v2}, 2014.

\bibitem[Giacobino et~al.(2016)Giacobino, Sardy, and Hengartner]{Sardy16}
C.~Giacobino, S.~Sardy, and N.~Hengartner.
\newblock Quantile universal threshold selection with an application in
  generalized model with lasso.
\newblock Technical report, arXiv, Cornell, 2016.

\bibitem[Goodman(1974)]{Good74}
L.~A. Goodman.
\newblock Exploratory latent structure analysis using both identifiable and
  unidentifiable models.
\newblock \emph{Biometrika}, 61:\penalty0 255--231, 1974.

\bibitem[Greenacre(1984)]{Greenacre84}
M.~J. Greenacre.
\newblock \emph{Theory and Applications of Correspondence Analysis}.
\newblock Acadamic Press, 1984.

\bibitem[Greenacre and Blasius(2006)]{Green06}
M.~J. Greenacre and J.~Blasius.
\newblock \emph{Multiple Correspondence Analysis and Related Methods}.
\newblock Chapman \& Hall/CRC, 2006.

\bibitem[Josse and Sardy(2015)]{Josse15}
J.~Josse and S.~Sardy.
\newblock Adaptive shrinkage of singular values.
\newblock \emph{Statistics and Computing}, pages 1--10, 2015.

\bibitem[Kiers(1997)]{Kiers97}
H.~A.~L. Kiers.
\newblock Weighted least squares fitting using ordinary least squares
  algorithms.
\newblock \emph{Psychometrika}, 62\penalty0 (2):\penalty0 251--266, 1997.

\bibitem[Lange(2004)]{Lang04}
K.~Lange.
\newblock \emph{Optimization}.
\newblock Springer-Verlag, New York, 2004.

\bibitem[le~Roux(2010)]{leroux10}
B.~le~Roux.
\newblock \emph{Multiple Correspondence Analysis}.
\newblock SAGE publications, CA: Thousand Oaks, 2010.

\bibitem[Li and Tao(2013)]{Li2013SEPCA}
J.~Li and D.~Tao.
\newblock Simple exponential family {PCA}.
\newblock \emph{Neural Networks and Learning Systems, IEEE Transactions on},
  24\penalty0 (3):\penalty0 485--497, 2013.

\bibitem[Little and Rubin(1987, 2002)]{Little02}
R.~J.~A. Little and D.~B. Rubin.
\newblock \emph{Statistical Analysis with Missing Data}.
\newblock Wiley series in probability and statistics, New-York, 1987, 2002.

\bibitem[Michailidis and De~Leeuw(1998)]{Mich98}
G.~Michailidis and J.~De~Leeuw.
\newblock The {Gifi} system of descriptive multivariate analysis.
\newblock \emph{Statistical Science}, 13:\penalty0 307--336, 1998.

\bibitem[Nishisato(1980)]{Nishisato80}
S~Nishisato.
\newblock \emph{Analysis of Categorical Data: Dual Scaling and its
  Applications}.
\newblock University of Toronto Press, Toronto, 1980.

\bibitem[Rubin(1987)]{Rubin87}
D.~B. Rubin.
\newblock \emph{Multiple Imputation for Non-Response in Survey}.
\newblock Wiley, 1987.

\bibitem[Shabalin and Nobel(2013)]{Shabalin2013}
Andrey~A Shabalin and Andrew~B Nobel.
\newblock Reconstruction of a low-rank matrix in the presence of {G}aussian
  noise.
\newblock \emph{Journal of Multivariate Analysis}, 118:\penalty0 67--76, 2013.

\bibitem[Srebro(2004)]{Srebro04}
N~Srebro.
\newblock \emph{Learning with Matrix Factorizations}.
\newblock PhD thesis, Massachusetts institute of technology, 2004.

\bibitem[Tenenhaus and Young(1985)]{Tenenhaus85}
M.~Tenenhaus and F.~W. Young.
\newblock An analysis and synthesis of multiple correspondence analysis,
  optimal scaling, dual scaling, homogeneity analysis and other methods for
  quantifying categorical multivariate data.
\newblock \emph{Psychometrika}, 50:\penalty0 91--119, 1985.

\bibitem[Tibshirani(1996)]{Tibs:regr:1996}
R.~Tibshirani.
\newblock Regression shrinkage and selection via the {Lasso}.
\newblock \emph{Journal of the Royal Statistical Society, Series B:
  Methodological}, 58:\penalty0 267--288, 1996.

\bibitem[Tibshirani and Taylor(2012)]{Tib12}
R.~J. Tibshirani and J.~Taylor.
\newblock Degrees of freedom in lasso problems.
\newblock \emph{The Annals of Statistics}, 40 (2):\penalty0 1198--1232, 2012.

\bibitem[{Van Buuren}(2012)]{VB12}
S.~{Van Buuren}.
\newblock \emph{Flexible Imputation of Missing Data (Chapman \& Hall/{CRC}
  Interdisciplinary Statistics)}.
\newblock Chapman and Hall/CRC, 2012.

\bibitem[Zou et~al.(2007)Zou, Hastie, and Tibshirani]{Zou07}
H.~Zou, T.~Hastie, and R.~Tibshirani.
\newblock On the "degrees of freedom" of the lasso.
\newblock \emph{The Annals of Statistics}, 35 (2):\penalty0 2173--2192, 2007.

\end{thebibliography}

\end{document}